\documentclass[aps,nofootinbib,prd,eqsecnum,showpacs,showkeys,preprintnumbers]{revtex4-2}
\usepackage{graphicx}
\usepackage{wrapfig}
\usepackage{caption}
\usepackage{subcaption}
\usepackage{amsmath}
\usepackage{amsfonts}
\usepackage{amssymb}
\usepackage{color}
\usepackage{bm}
\usepackage{natbib}
\usepackage{epstopdf}
\usepackage{url}
\usepackage{footnote}
\usepackage{textcomp}
\usepackage{ulem}
\usepackage{esint}
\usepackage[unicode=true, pdfusetitle,
 bookmarks=true,bookmarksnumbered=false,bookmarksopen=false,
 breaklinks=false,pdfborder={0 0 1},backref=false,colorlinks=false]{hyperref}
\usepackage{multirow}
\usepackage{pifont}
\usepackage{lipsum}
\usepackage[mathlines]{lineno}
\usepackage[newcommands]{ragged2e}
\begin{document}
\title{Gravitational waves production during preheating within GB gravity \\with monomial coupling}
\author{Brahim Asfour}
\email{brahim.asfour@ump.ac.ma}
\author{Yahya Ladghami}
\email{yahyaladghami@gmail.com}
\author{Taoufik Ouali}
\email{t.ouali@ump.ac.ma}
\affiliation{Laboratory of Physics of Matter and Radiation, \\
University of Mohammed first, BP 717, Oujda, Morocco\\
Astrophysical and Cosmological Center, Faculty of Sciences, BP 717, Oujda, Morocco}

\date{\today }
\begin{abstract}
In this paper, we investigate the production of gravitational waves during the preheating era. To achieve this purpose, we consider Gauss-Bonnet inflation model with Power{\textendash}law potential, $V(\phi)= V_0 \phi^n$, and monomial Gauss-Bonnet coupling function, $\xi(\phi)= \xi_0 \phi^n$. We examine our model by comparing our findings with the current observational data. After that, we study the preheating stage by adopting an approach in which we establish a link between preheating duration, reheating phase and inflationary parameters. This step allows us to benefit from observational constraints imposed on inflation.  Furthermore, we examine the production of gravitational waves during preheating epoch connecting the energy density to the preheating duration, $N_{pre}$, and then with the spectral index $n_s$. The generation of gravitational waves during preheating can satisfy observational constraints. In particular, the predicted present-day gravitational-wave energy density, expressed as a function of the scalar spectral index, is consistent with the Planck constraints for the choice of a dimensionless Gauss-Bonnet coupling parameter $\alpha \equiv 4V_{0}\xi_{0}/3 = -1.5\times 10^{-6}$, an effective equation of state parameter $\omega = 1/6$, and a preheating efficiency parameter $\delta = 10^{5}$.

\end{abstract}
\maketitle
\section{Introduction}
Inflation has become the leading framework to study the early Universe and resolve the shortcomings of the standard Big Bang model, namely the flatness and horizon problems. Additionally, inflation explains the origin of the large-scale structures observed today. According to the inflationary paradigm, the Universe underwent a brief phase of rapid expansion during its earliest stages. The single scalar field assumed to be responsible for this accelerated expansion is referred to as inflaton \cite{Starobinsky:1980te, Guth:1980zm, Linde:1981mu, Albrecht:1982wi, Linde:1983gd, Kazanas:1980tx, Bargach:2019pst, Bargach:2020bpf, Asfour:2022qap,Bouabdallaoui:2016izz,Bouabdallaoui:2022wyp}. After inflation, the Universe is left cold because of this kind of expansion. Therefore, it is necessary to introduce an intermediate phase between inflation and the radiation-dominated era, known as reheating \cite{Abbott:1982hn, Dolgov:1982th, Albrecht:1982mp, Traschen:1990sw}. This phase serves to heat and thermalize the Universe by converting the energy stored in the inflaton field into particles \cite{Shtanov:1994ce, Kofman:1994rk, Amin:2014eta, Asfour:2024mfr, Bargach:2025ppb,Kofman:1997yn}. However, this mechanism can be slow and inefficient. Consequently, a preheating stage is incorporated as the onset of the reheating period. Preheating is an explosive, non-perturbative process that occurs immediately after inflation, during which the energy density of the inflaton is rapidly transferred to coupled fields through parametric resonance, producing a large number of particles \cite{Greene:1997fu, Dolgov:1989us}.
\par To constrain this post-inflationary epoch, it is necessary to relate its parameters, specifically the preheating duration, to those of reheating and inflation, such as the reheating duration, thermalization temperature, and spectral index. This approach enables us to utilize the recent observational data provided by Planck experiments \cite{Planck:2018jri}. Furthermore, the preheating phase is characterized by the equation-of-state (EoS) parameter, $\omega$, which describes the cosmic fluid and is defined as $\omega = p/\rho$. In our analysis, we assume that this EoS remains constant during this era, satisfying the following interval $-1/3 \leq \omega \leq 1$. The lower bound ensures that the reheating expansion is not accelerated~\cite{Dai:2014jja}, while the upper bound is required to preserve causality. We also introduce an additional key parameter, $\delta$, which represents the efficiency of the preheating process relating the energy density at the end of inflation, $\rho_{\rm end}$, to that at the end of preheating, $\rho_{\rm pre}$ \cite{ElBourakadi:2021nyb}. 
\par Moreover, the explosive production of particles during preheating produce large and significant inhomogeneity, which can act as an effective source of gravitational waves (GW) with large energy density \cite{Khlebnikov:1997di, Garcia-Bellido:2008ycs, Sfakianakis:2018lzf, DeCross:2016cbs, DeCross:2016fdz, Ema:2016dny, Repond:2016sol, DeCross:2015uza,Felder:2001kt, Felder:2000hj,Rubio:2019ypq, Dux:2022kuk, ElBourakadi:2022anr, Oikonomou:2023qfz, Oikonomou:2024zhs,ElBourakadi:2021nyb,ElBourakadi:2022lqf}. According to general relativity (GR), GW are ripples in the curvature of space-time generated in the early Universe and propagate at the speed of light. Consequently, the detection of these gravitational waves would provide a powerful probe of inflationary models and of the preheating stage \cite{Aggarwal:2020olq,Franciolini:2022htd}. In our study, we focus on the current energy density of gravitational waves. We adopt an approach where we correlate the current GW energy density with the key parameters of preheating and inflation.
\par On the other hand, Gauss-Bonnet (GB) theory is one of the most successful alternatives to General Relativity. It plays a significant role in addressing cosmological issues and enhances our understanding of the early Universe. Numerous studies have investigated Gauss–Bonnet gravity in both the four and higher dimensions \cite{Satoh:2010ep, Guo:2010jr, Jiang:2013gza, Townsend:1979js, Boulware:1985wk,Koh:2014bka}. Commonly, researchers derive the main inflationary parameters, such as the scalar spectral index $n_s$ and the tensor-to-scalar ratio $r$, and check the viability of the theoretical predictions by considering a specific model with a power-law potential and an inverse monomial Gauss-Bonnet (GB) coupling function \cite{Guo:2010jr,Jiang:2013gza}. In Refs. \cite{Koh:2018qcy,ElBourakadi:2021nyb}, the authors used this model to study and set observational constraints on primordial gravitational waves. However, inverse monomial coupling satisfying the relation $V(\phi)\xi(\phi)=$constant, may not lead to the oscillations of the inflaton field around the minimum of its potential, which are required for the presence of the non-perturbative preheating \cite{Koh:2014bka}. Consequently, to overcome this issue, we employ a power-law potential alongside a monomial GB coupling function, as a framework to examine the preheating stage and explore the production of gravitational waves during this epoch.
\par The paper is structured as follows: In section II, we briefly review the main equations and parameters of the Gauss-Bonnet inflationary model and compare our obtained findings with the recent Planck data \cite{Planck:2018jri}. In section III, we study the preheating phase, deriving the preheating duration and relating it to the main parameters of inflation and reheating. In section IV, we focus on the generation of primordial gravitational waves during preheating period by establishing a link between GW, inflation and preheating parameters. Finally, we discuss and conclude our results in section V.

\section{Slow-roll GB inflation}
Let us consider the following action describing the Gauss–Bonnet model, which includes both the Einstein–Hilbert and Gauss–Bonnet terms coupled to the scalar field $\phi$ \cite{Satoh:2010ep,Guo:2010jr,Koh:2014bka,Jiang:2013gza,Koh:2016abf}
\begin{equation}
S=\int d^4x \sqrt{-g}\left[\frac{1}{2\kappa^{2}}R-\frac{1}{2} g^{\mu \nu} \nabla_{\mu} \phi \nabla_{\nu} \phi - V(\phi)- \frac{1}{2}\xi(\phi)R^{2}_{GB}\right],
\end{equation}
where $R^{2}_{GB}=R^{2}-4R_{\mu\nu}R^{\mu\nu}+R_{\mu\nu\rho\sigma}$ denotes the GB term, $\xi(\phi)$ represents the coupling function and $\kappa^{2}=M^{-2}_{p}$ is the reduced Planck mass. The corresponding background dynamical equations in a Friedmann-Robertson-Walker Universe are given by \cite{Koh:2018qcy}
\begin{equation}
H^2=\frac{\kappa^{2}}{3}\left(\frac{1}{2}\dot{\phi}^2+V+12\dot{\xi}H^{3}\right),\\   \label{eq2.2}
\end{equation}
\begin{equation}
\dot{H}=-\frac{\kappa^{2}}{2}\left[\dot{\phi}^2-4\ddot{\xi}H^2-4\dot{\xi}H\left(2\dot{H}-H^2\right)\right], \label{eq2.3}
\end{equation}
and
\begin{equation}
\ddot{\phi}+3H\dot{\phi}+V_{\phi}+12\xi_{\phi}H^2\left(\dot{H}+H^2\right)=0, \label{eq2.4}
\end{equation}
where the overdot denotes differentiation with respect to cosmic time, H is the Hubble parameter, and $V_{\phi}$ and $\xi_{\phi}$ indicate the derivative of the potential $V(\phi)$ and the coupling function $\xi(\phi)$ with respect to scalar field $\phi$, respectively.\linebreak

Under the slow-roll approximation and the conditions associated with the Gauss–Bonnet coupling
\begin{eqnarray}
\dot{\phi}^2/2 \ll  V, \hspace{1cm}  \ddot{\phi}\ll 3H\dot{\phi},\hspace{1cm} 4H \dot{\xi} \ll 1 ,\hspace{1cm} \ddot{\xi}\ll \dot{\xi}H, 
\end{eqnarray}
Eqs. \eqref{eq2.2} and \eqref{eq2.3} can be rewritten in the following reduced forms 

\begin{eqnarray}
H^2&=&\frac{\kappa^{2}}{3}V,   \label{eq2.6}\\ 
\dot{H}&=&-\frac{\kappa^{2}}{2}\left[\dot{\phi}^2+4\dot{\xi}H^3\right],
\end{eqnarray}
and
\begin{equation}
3H\dot{\phi}+V_{\phi}+12\xi_{\phi}H^4=0.
\end{equation}
The slow-roll parameters can be expressed in terms of the potential and the coupling function as \cite{Koh:2018qcy}
\begin{eqnarray}
&\epsilon=&\frac{Q}{2\kappa^2}\frac{V_{\phi}}{V}, \label{eq2.9}\\
&\eta=&-\frac{Q}{\kappa^2}\left( \frac{V_{\phi \phi}}{V_{\phi}}+\frac{Q_{\phi}}{Q}\right) ,\\
&\delta_1=&-\frac{4\kappa^2}{3}\xi_{\phi}Q V,
\end{eqnarray}
and
\begin{eqnarray}
&\delta_2=&-\frac{Q}{\kappa^2}\left( \frac{\xi_{\phi \phi}}{\xi_{\phi}}+\frac{1}{2}\frac{V_{\phi}}{V}+\frac{Q_{\phi}}{Q}\right),
\end{eqnarray}
where 
\begin{eqnarray}
Q=\frac{V_{\phi}}{V}+\frac{4}{3}\kappa^4 \xi_{\phi} V.
\end{eqnarray}
The number of e-folds $N$ is given by 
\begin{eqnarray}
N_k=\int_{t_k}^{t_{end}}Hdt= \int_{\phi_{end}}^{\phi_k}\frac{\kappa^{2}}{Q}d\phi, \label{eq2.14}
\end{eqnarray}
where "$k$" and "$end$" indicate the time of the horizon crossing and the end of inflation, respectively.
\linebreak
The spectral index and the tensor-to-scalar ratio are given in terms of slow-roll parameters by \cite{Koh:2018qcy}
\begin{equation}
n_s-1=-2\epsilon-\frac{2\epsilon(2\epsilon+\eta)-\delta_1(\delta_2-\epsilon)}{2\epsilon-\delta_1}, \label{eq2.15} 
\end{equation}
and
\begin{equation}
r=8(2\epsilon-\delta_1) \label{eq2.16}.
\end{equation}

To illustrate our purpose, we examine the Gauss–Bonnet model, we consider the following power-law potential and monomial coupling function \cite{Koh:2014bka}
\begin{eqnarray}
V(\phi)= V_0 \phi^n, \hspace{1cm} \xi(\phi)= \xi_0 \phi^n,
\end{eqnarray}
where $V_{0}$ and $\xi_{0}$ are dimensionless constants and $n$ is taken to be positive. Using Eqs. \eqref{eq2.14}-\eqref{eq2.16}, the inflationary observables, namely the spectral index and tensor-to-scalar ratio, can be written as functions of the e-folding number $N_k$ as follows
\begin{eqnarray}
n_s-1=\frac{(n+2)}{2 N_k}+\frac{n(3n+2)(2n N_k)^n \alpha}{2(1+n)N_k \kappa^{2n}}, 
\end{eqnarray}
\begin{eqnarray}
r=\frac{4n}{N_k} + \frac{4n(2n+1)(2n N_k)^n \alpha}{(1+n)N_k \kappa^{2n}},
\end{eqnarray}
where $\alpha=4V_{0}\xi_0/3$  .
\captionsetup[figure]{justification=Justified}
\begin{figure}[h!]
\centering
\begin{subfigure}[b]{0.49\textwidth}
\includegraphics[scale=0.55]{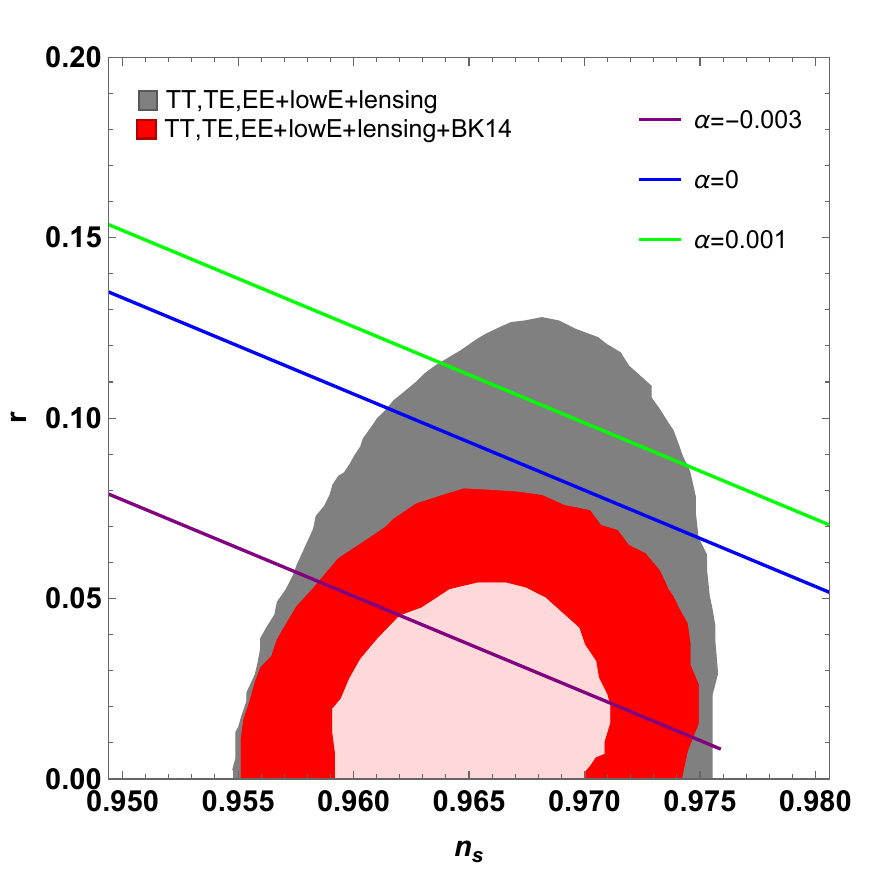} 
\caption*{-1a-}
\end{subfigure}
\begin{subfigure}[b]{0.49\textwidth}
\includegraphics[scale=0.55]{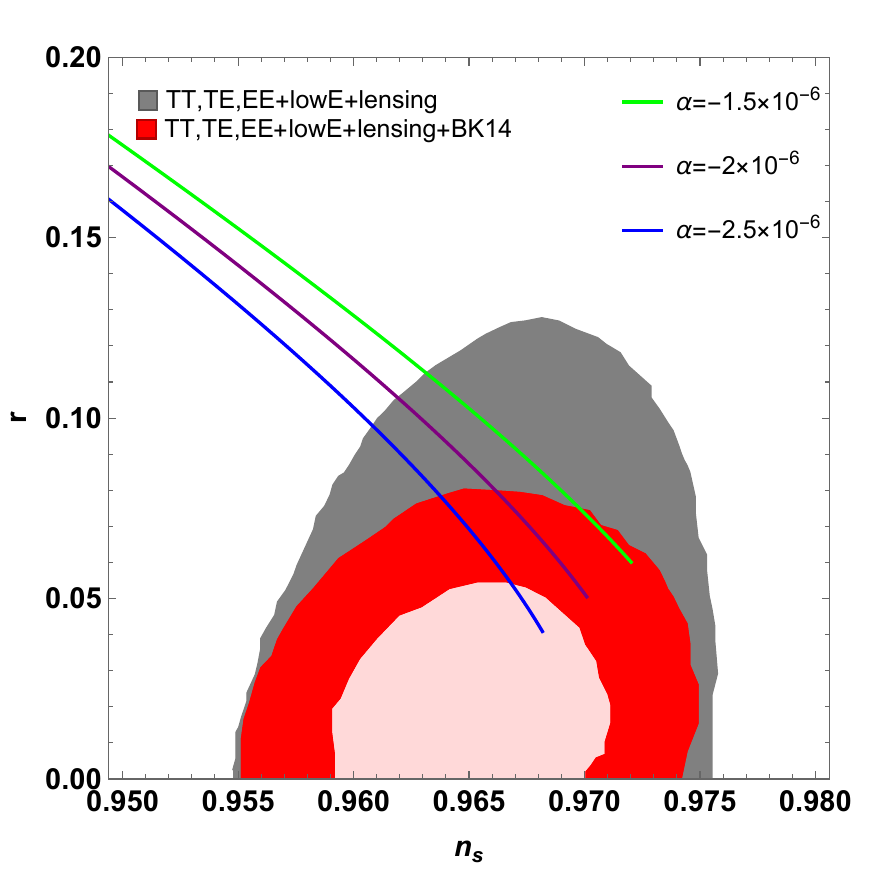} 
\caption*{-1b-}
\end{subfigure} 
\caption{Tensor-to-scalar ratio, $r$, as a function of the spectral index $n_s$ considering three values of $\alpha$, for $n=1$ (left panel) and $n=2$ (right panel). The red and gray contours indicate $1\sigma$ and $2\sigma$ constraints from Planck, respectively. } 
\label{Fig:1}
\end{figure}
Fig.\ref{Fig:1} illustrates the evolution of the tensor-to-scalar ratio as a function of the spectral index considering three distinct values of $\alpha$. The left panel depicts the case $n=1$, whereas the right panel corresponds to $n=2$. We also include the constraints from the Planck TT,TE,EE+LowE+lensing (gray contour) and Planck TT,TE, EE+lowE+lensing+BK14 data (red contour) \cite{Planck:2018jri}. The results demonstrate that theoretical predictions for both values of $n$ are consistent with observational data for all considered values of $\alpha$. Furthermore, the tensor-to-scalar ratio increases as the parameter $\alpha$ increases.
\section{Preheating stage}
To extract information about preheating phase, we consider different eras constituting the history of the Universe between the horizon crossing and the present time. By setting $k=a_kH_k$ at the horizon crossing, this cosmic process could be described by the following expression
\begin{equation}
\frac{k}{a_{0}H_{0}}=\frac{a_k}{a_{end}}\frac{a_{end}}{a_{pre}}\frac{a_{pre}}{a_{re}}\frac{a_{re}}{a_0}\frac{H_k}{H_{0}}, \label{eq3.1}
\end{equation}
where $a_{0}$, $a_k$, $a_{pre}$ $a_{re}$, and $a_{end}$ indicate the scale factor at the present time, at the horizon crossing time, at the end of preheating, at the end of reheating, and at the end of inflation, respectively. Whereas $H_{0}$ and $H_{k}$ are the Hubble constant at the present and at the crossing time, respectively. Introducing the number of e-folds during inflation, $N_k=\ln(a_{end}/a_k)$, the number of e-folds between the end of inflation and the end of preheating, $N_{pre}=\ln\left(a_{pre}/a_{end} \right)$, and the number of e-folds between the end of preheating and the end of reheating, $N_{re}=\ln(a_{re}/a_{pre})$, the above equation can then be rewritten as
\begin{equation}
\ln\left(\frac{k}{a_{0}H_{0}}\right)= -N_k-N_{pre}-N_{re}+\ln\frac{a_{re}}{a_0}+\ln\frac{H_k}{H_{0}}. \label{eq3.2}
\end{equation}
We assume that no entropy production takes place after the Universe reaches the thermalization temperature. Using the entropy conservation between reheating and present time, we can write \cite{Koh:2018qcy}
\begin{eqnarray}
\frac{a_{re}}{a_0}=\frac{T_0}{T_{re}}\left(\frac{43}{11 g_{re}}\right)^{1/3}, \label{eq3.3}
\end{eqnarray} 
Moreover, $g_{re}=g(T_{re})$ represents the number of relativistic degrees of freedom at the end of reheating characterized by the reheating temperature, $T_{re}$, and $T_{0}=2.725 K$ is the present CMB temperature. The reheating energy density is related to the thermalization temperature through the following expression
\begin{equation}
\rho_{re}=\frac{\pi^2}{30}g_{re} T_{re}^4. \label{eq3.4}
\end{equation}
In addition, we consider that the energy density at the end of inflation is connected to the preheating energy density by the relation \cite{ElBourakadi:2021nyb}
\begin{equation}
\rho_{end}= \delta \rho_{pre},
\end{equation}
where the parameter $\delta$ quantifies the efficiency of energy transfer from the inflaton field to other fields during the preheating phase. Using the relation $\rho\propto a^{-3(1+\omega)}$ to relate the energy density during the preheating phase to that of the reheating phase, where the equation of state parameter, $\omega=p/\rho$, is assumed to remain constant and lie within the range $-\frac{1}{3}\leq\omega\leq1$, we obtain
\begin{eqnarray}
\nonumber \rho_{re}&=&\frac{\rho_{end}}{\delta}(\frac{a_{re}}{a_{pre}})^{-3(1+\omega)}\\
&=&\frac{\rho_{end}}{\delta} e^{-3(1+\omega)N_{re}}, \label{eq3.6} 
\end{eqnarray}
where the number of e-folds of reheating, $N_{re}$, was introduced. Furthermore, the energy density at the end of inflation, $\rho_{end}$, may be written in terms of the potential $V_{end}$ as \cite{Koh:2018qcy}
\begin{equation}
\rho_{end}=\lambda_{end}V_{end}, \label{eq3.7} 
\end{equation}
where $\lambda_{end}$ is the effective ratio of kinetic energy to potential energy at the end of inflation, which can be  obtained from Eq.\eqref{eq2.2} as follows
\begin{equation}
\lambda_{end}=\left[\frac{1}{2}\frac{\dot{\phi}^2}{V(\phi)}+1+\frac{12\dot{\xi}H}{V(\phi)} \right]_{\phi=\phi_{end}}.
\end{equation} 
At the end of inflation, the first slow-roll parameter satisfies the condition $\epsilon(\phi_{end})=1$. From Eq. \eqref{eq2.9} we obtain the following expression  
\begin{equation}
\phi_{end}=\left[  \frac{\frac{\kappa^2}{2}-\sqrt{\frac{\kappa^4}{4}-4\alpha}}{2\alpha}\right]^{1/2},
\end{equation}
and the potential at the end of inflation takes the form
\begin{equation}
V_{end}=V_{0}\left[\frac{\frac{\kappa^2}{2}-\sqrt{\frac{\kappa^4}{4}-4\alpha}}{2\alpha}\right]^{n/2}.
\end{equation}
To express the preheating duration in terms of reheating e-folds number and inflationary quantities, we use Eqs. \eqref{eq3.3}, \eqref{eq3.4}, \eqref{eq3.6} and \eqref{eq3.7}. Hence we obtain the following expression
\begin{eqnarray}
N_{pre}=\left[-\ln\left( \frac{k}{a_0 T_0}\right) -\frac{1}{3}\ln\left( \frac{11 g_{re}}{43}\right)-\frac{1}{4}\ln\left(\frac{30\lambda_{end}}{ \delta g_{re} \pi^2}\right)-\frac{1}{4}\ln\left(\frac{V_{end}}{H_{k}^4}\right)-N_k \right]-\frac{1-3\omega}{4}N_{re}, \label{eq3.9}
\end{eqnarray}
which can be simplified by considering the numerical values \cite{Planck:2018jri}: $a_{0}=1$, $k=0.05$Mpc$^{-1}$, $T_{0}=2.725$K and $g_{re}\approx106.75$, as
\begin{eqnarray}
N_{pre}=\left[60.0085-\frac{1}{4}\ln\left(\frac{30\lambda_{end}}{ \delta g_{re} \pi^2}\right)-\frac{1}{4}\ln\left(\frac{V_{end}}{H_{k}^4}\right)-N_k \right]-\frac{1-3\omega}{4}N_{re}. \label{eq3.10}
\end{eqnarray}
Furthermore, under the slow-roll assumption, the Hubble parameter at the time of horizon crossing can be expressed as follows 
\begin{equation}
H_k=\left( \frac{\kappa^2 V_0}{3} \right)^{1/2}\left[\sqrt{\frac{2 n N_k}{\kappa^2}}\left( 1 + \frac{\alpha (2 n N_k)^n}{2 (n + 1) \kappa^{2n}} \right)\right]^{n/2},
\end{equation}
where we have used the expression of the inflaton field derived from Eq.\eqref{eq2.14} 
\begin{equation}
\kappa\phi_k= \sqrt{2 n N_k}\left(1 + \frac{\alpha (2 n N_k)^n}{2 (n + 1) \kappa^{2n}}\right).
\end{equation}
In our calculation, the large field inflation scenario with the approximation $\phi_k\geq \phi_{end}$ is adopted. Additionally, from Eqs. \eqref{eq3.4} and \eqref{eq3.7}, the reheating duration can be expressed as a function of thermalization temperature as follows
\begin{equation}
N_{re}=\frac{1}{3(1+\omega)}\ln\left(\frac{30\lambda_{end}V_{end}}{\delta \pi^2 g_{re}T_{re}^4}\right).
\end{equation}
\captionsetup[figure]{justification=Justified}
\begin{figure}[h!]
\centering
\includegraphics[scale=0.55]{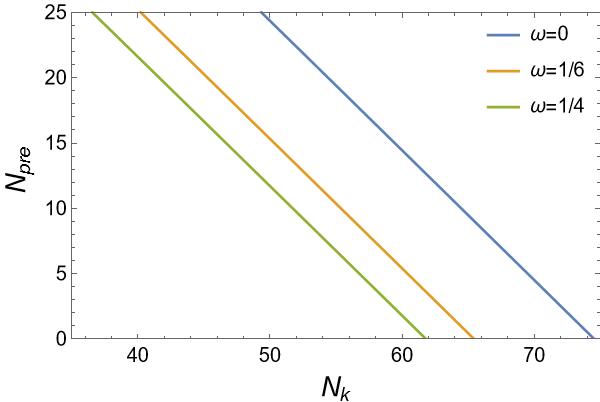} 
\caption{Preheating duration, $N_{pre}$, against the number of e-folds, $N_k$, considering three values of $\omega$, for $\alpha=-1.5\times10^{-6}$.} 
\label{Fig:2}
\end{figure}
\par Fig.~\ref{Fig:2} illustrates the evolution of the preheating duration, \(N_{\text{pre}}\), as a function of the number of e-folds, \(N_k\), for three different values of the equation of state parameter: \(\omega = 0,\, 1/6,\, 1/4\), and for the coupling parameter \(\alpha = -1.5 \times 10^{-6}\). Our findings reveal that the preheating duration is sensitive to EoS. Notably, \(\omega = 1/4\) leads to the shortest preheating phase, as \(N_{\text{pre}}\) is smaller, indicating the most efficient preheating process.
\captionsetup[figure]{justification=Justified}
\begin{figure}[h!]
\centering
\begin{subfigure}[b]{0.49\textwidth}
\includegraphics[scale=0.55]{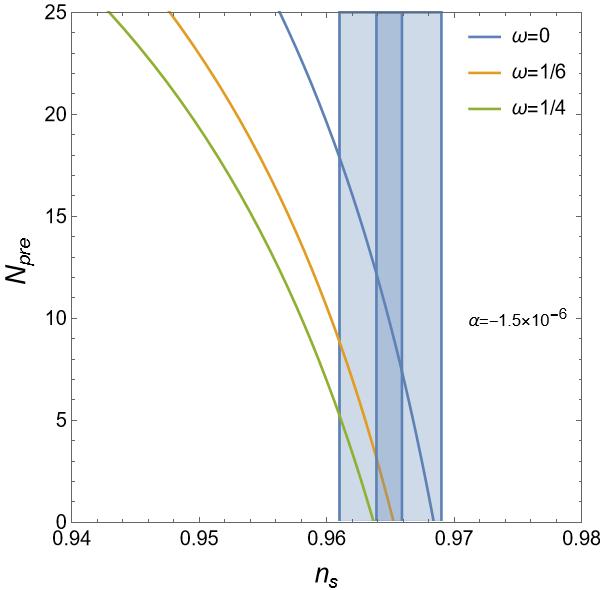} 
\caption*{-3a-}
\end{subfigure}
\begin{subfigure}[b]{0.49\textwidth}
\includegraphics[scale=0.55]{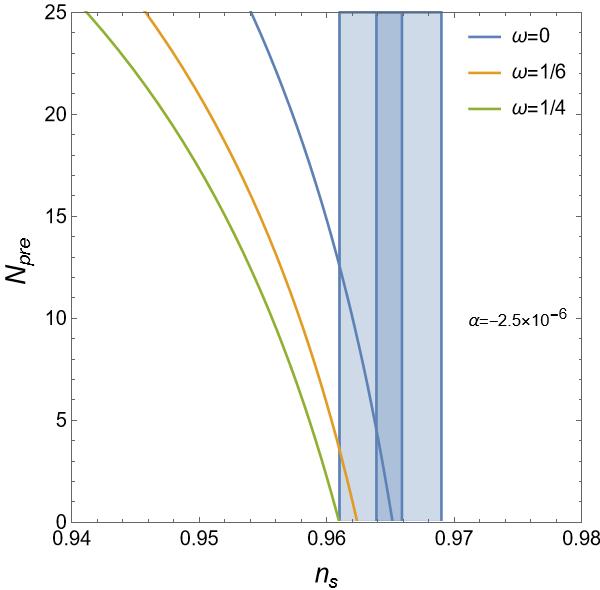} 
\caption*{-3b-}
\end{subfigure} 
\caption{Preheating duration $N_{pre}$ against the spectral index $n_s$ considering three values of $\omega$, for $\alpha=-1.5\times10^{-6}$ (left panel) and $\alpha=-2.5\times10^{-6}$ (right panel). The vertical light blue band represents the Planck constraints on the scalar spectral index, $n_s=0.9649\pm 0.0042$, whereas the dark blue band indicates the anticipated $10^{-3}$ precision expected from future experiments \cite{Amendola:2016saw}.} 
\label{Fig:3}
\end{figure}
\par To benefit from observational constraints on inflationary parameters, we plot the preheating duration, $N_{pre}$, as a function of the spectral index, $n_s$, as shown in Fig. \ref{Fig:3}. We consider three values of the equation of state, $\omega=0, 1/6$ and $1/4$ assuming a reheating temperature of $T_{re}=10^{16}$ GeV and a preheating parameter value of  $\delta=10^{5}$. The vertical light blue region corresponds to the Planck bounds on $n_s=0.9649\pm 0.0042$, while the dark blue region represents precision of $10^{-3}$ from future experiments \cite{Amendola:2016saw}.  We find that all curves converge to the preheating instant point corresponding to $N_{pre}\rightarrow0$. As all curves for both $\alpha=-1.5\times10^{-6}$ and $\alpha=-2.5\times10^{-6}$ lie within the Planck bounds \cite{Planck:2018jri}. We conclude that all selected values of $\omega$ are consistent with current observational data.
\section{Gravitational waves production}
The intense and rapid production of particles during the preheating stage may induce significant metric fluctuations, which can serve as a source of gravitational waves. In this section, we investigate the generation of GW within preheating phase. In a flat Friedmann–Robertson–Walker (FRW) background, gravitational waves can be described by the traceless part of the spatial metric perturbations \cite{Dufaux:2007pt,Adshead:2018doq}
\begin{equation}
ds^2=-dt^2+a^2(t)[(\delta_{ij}+h_{ij})dx^idx^j],
\end{equation}
where the transverse-traceless conditions, $\partial_ih_{ij}=0$ and $h_{ii}=0$, are satisfied. The evolution of the transverse-traceless tensor perturbations $h_{ij}$ is described be the following equation of motion \cite{Li:2020qnk} 
\begin{equation}
\ddot{h}_{ij}+3H\dot{h}_{ij}-\frac{1}{a^2}\nabla^2h_{ij}=16\pi^2G\Pi_{ij}^{TT}.
\end{equation}
The source term $\Pi_{ij}^{TT}$ is the transverse-traceless part of the full anisotropic-stress energy tensor, $\Pi_{ij}$, given by 
\begin{equation}
a^2\Pi_{ij}=T_{ij}-\langle p\rangle g_{ij},
\end{equation}
where $\langle p\rangle$ is the background homogeneous pressure. The energy density carried by GW can be defined 
through the following equation \cite{Misner:1973prb}
\begin{equation}
\rho_{gw,0}=\frac{1}{32\pi G} \langle\dot{h}_{ij}(t,x)\dot{h}_{ij}(t,x)\rangle.
\end{equation} 
The current abundance of the gravitational waves energy density is given by
\begin{equation}
h^2\left(\frac{\rho_{gw,0} }{\rho_{c,0}} \right)=\int\frac{df}{f} h^2 \Omega_{gw,0}(f),
\end{equation}
and the GW energy spectrum can be written as 
\begin{equation}
\Omega_{gw,0}(f) h^2=\frac{h^2}{\rho_{c,0}}\frac{d \rho_{gw,0} }{d  \ln f},
\end{equation}
where $\rho_{c,0}=3H_0^2/(8\pi G)$ is the current critical energy density and $f$ is the frequency. To study the GW energy density spectrum, we need to relate it to the current physical quantities constrained by observations. The ratio of the current scale factor to the one at the end of inflation is expressed as \cite{Dufaux:2007pt,Adshead:2018doq} 
\begin{equation}
\frac{a_{end}}{a_0}=\frac{a_{end}}{a_{pre}}\left(\frac{a_{pre}}{a_{re}}\right)^{1-\frac{3}{4}(1+\omega)}\left(\frac{g_{re}}{g_{0}}\right)^{-1/12}\left(\frac{\rho_{rad,0}}{\rho_{pre}}\right)^{1/4}, \label{eq4.7}
\end{equation}
where the GW production is assumed to stop at the end of preheating. $\rho_{rad,0}$ indicates the present energy density of radiation and $g_{re}/g_{0}=106.75/3.36\simeq 31$. The corresponding physical frequency today is defined as \cite{Li:2020qnk}
\begin{equation}
f=\frac{k}{2\pi a_0}=\frac{k}{a_{pre}\rho_{pre}^{1/4}}\left(\frac{a_{pre}}{a_{re}}\right)^{1-\frac{3}{4}(1+\omega)}\times (4\times 10^{10} Hz).
\end{equation}
In our analysis, we focus on the current energy density of gravitational waves, so we connect this quantity to the GW spectra generated during the preheating era. The GW energy spectrum, with the help of Eq. \eqref{eq4.7} and the fact that GW behaves like radiation with the cosmic expansion, writes \cite{ElBourakadi:2022lqf,Li:2020qnk} 
\begin{equation}
\Omega_{gw,0}(f)h^2=\Omega_{gw}(f) \delta\left(\frac{a_{end}}{a_{pre}}\right)^4 \left(\frac{a_{pre}}{a_{re}}\right)^{(1-3\omega)} \left( \frac{g_{re}}{g_{0}} \right)^{-1/3} \Omega_{r,0}h^2,
\end{equation}
Employing the relations $a_{end}/a_{pre}=e^{-N_{pre}}$ and $a_{pre}/a_{re}=e^{-N_{re}}$ together with Eq. \eqref{eq3.9}, we can rewrite the current gravitational wave energy density spectrum as follows
\begin{equation}
\Omega_{gw,0}h^2= \Omega_{gw}(f) \delta \left( \frac{g_{re}}{g_{0}} \right)^{-1/3} \Omega_{r,0}h^2  \exp\left\lbrace -4 \left[60.0085-\frac{1}{4}ln\left(\frac{30\lambda_{end}}{100 \delta \pi^2}\right)-\frac{1}{4}ln\left(\frac{V_{end}}{H_{k}^4}\right)-N_k \right]\right\rbrace.
\end{equation}
According to the recent Planck observational constraints, the current GW energy density spectrum is required to satisfy the upper bound $\Omega_{gw,0}h^2\leq 1.86\times 10^{-6}$ \cite{Planck:2018jri}.\\

\begin{figure}[h!]
\centering
\includegraphics[scale=0.55]{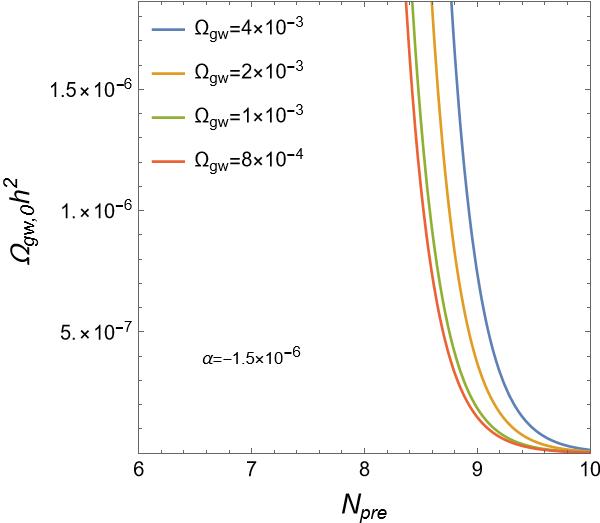}  
\caption{Variation of GW energy density versus the preheating duration, $N_{pre}$, for different values of the GW density spectra i.e. $ 4\times 10^{-3}$ (blue curve), $2\times 10^{-3}$ (orange curve), $10^{-3}$ (green curve) and $8\times 10^{-4}$ (red curve), with the coupling parameter $\alpha=-1.5\times10^{-6}$ and the equation of state $\omega=1/6$.} 
\label{Fig:4}
\end{figure}

The evolution of the current GW energy density as a function of the preheating duration is illustrated in Fig.\ref{Fig:4}. We examine four specific values of GW density spectrum, namely $\Omega_{gw}=$ $4\times 10^{-3}$, $2\times 10^{-3}$, $10^{-3}$ and $8\times 10^{-4}$. In our analysis, two specific values of the coupling parameter and the equation of state parameter are considered: $\alpha=-1.5\times10^{-6}$ and $\omega=1/6$. The results show good consistency with Planck observational constraints, as the current GW energy density remains below the upper bound, $\Omega_{gw,0}h^2\leq 1.86\times 10^{-6}$, for a suitable range of preheating e-folds number. Moreover, the GW density spectrum influence the preheating duration value, as  $N_{pre}$ increases with higher values of $\Omega_{gw}$.\par
\begin{figure}[h!]
\centering
\begin{subfigure}[b]{0.49\textwidth}
\includegraphics[scale=0.55]{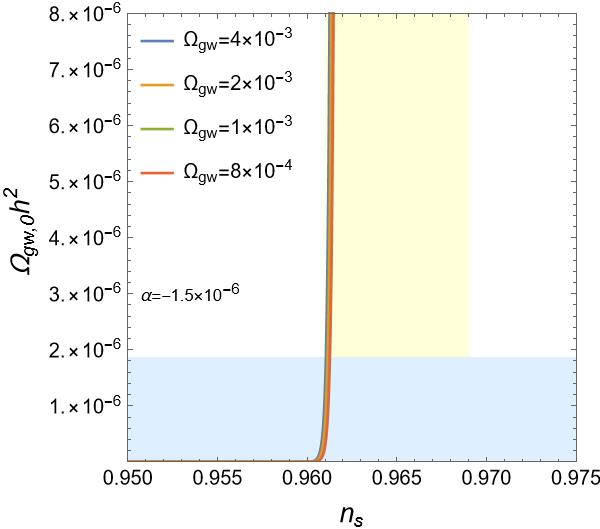} 
\caption*{-5a-}
\end{subfigure}
\begin{subfigure}[b]{0.49\textwidth}
\includegraphics[scale=0.55]{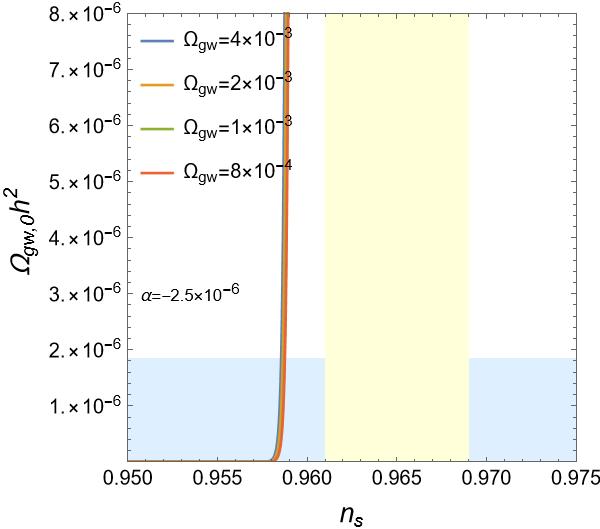} 
\caption*{-5b-}
\end{subfigure} 
\caption{Evolution of GW energy density as a function of the spectral index, $n_s$, for different values of the GW density spectra i.e. $ 4\times 10^{-3}$ (blue curve), $2\times 10^{-3}$ (orange curve), $10^{-3}$ (green curve) and $8\times 10^{-4}$ (red curve), considering two specific values of the coupling parameter $\alpha=-1.5\times10^{-6}$ (left plot) and $\alpha=-2.5\times10^{-6}$ (right plot). We take the equation of state $\omega=1/6$. The yellow vertical and blue horizontal regions indicate the Planck constraints on the spectral index, $n_s$, and the current GW energy density, $\Omega_{gw,0}h^2$, respectively.} 
\label{Fig:5}
\end{figure}

\captionsetup[figure]{justification=Justified}
In the following step, we establish a connection between the current GW energy density to the inflationary parameters, which allows us to use the recent Planck observational data imposed on inflation. We then investigate the current GW energy density behavior as a function of the spectral index, $n_s$, for two specific values of the coupling parameter $\alpha=-1.5\times10^{-6}$ and $\alpha=-2.5\times10^{-6}$ as shown in Fig.\ref{Fig:5}. In these plots, we fixed $\omega=1/6$ considering the same values of the GW density spectrum used in Fig.\ref{Fig:4}. Our findings indicate good agreement with observational data for $\alpha=-1.5\times10^{-6}$. For this value of $\alpha$, the current GW energy density satisfies the constraint $\Omega_{gw,0}h^2\leq 1.86\times 10^{-6}$ and all curves corresponding to GW density spectrum fall within the allowed bound of spectral index, $n_s$. In contrast, for $\alpha=-2.5\times10^{-6}$, the curves lie outside the range compatible with constraints imposed on $n_s$.

\captionsetup[figure]{justification=Justified}

\section{Conclusion}
In this paper, we have investigated the preheating stage occurring following the inflation period, during which the inflaton field oscillates and decays into daughter fields, resulting in the explosive production of particles. Moreover, we have examined the generation of primordial gravitational waves during preheating within the framework of Gauss-Bonnet gravity.
\par After reviewing the fundamental equations that describe GB inflation, we considered a power-law potential and a monomial coupling function. We computed the key inflation parameters such as the spectral index, and the tensor-to-scalar ratio, as a function of the number of e-folds. To test our theoretical predictions, we compared our findings to the recent observational data provided by the Planck collaboration.
\par To constrain the preheating phase, we adopt an approach in which we derive the preheating duration, in terms of inflation and reheating parameters. We then analyze the dependence of the preheating duration on the reheating duration, and the spectral index. To examine the impact of different parameters, we consider three distinct values of the equation of state and two specific values of the Gauss-Bonnet coupling. We found that the preheating duration is sensitive to the parameters $\alpha$ and the equation of state. 
\par Furthermore, we have reviewed the basic equations describing gravitational waves evolution. We focus on the present-day GW energy density relating this crucial parameter to the preheating duration and the spectral index. This method allows us to constrain the GW spectrum using the provided Observational bounds on inflation.
\par Consequently, we conclude that with an equation of state equal to $1/6$ and the coupling value of $\alpha = -1.5\times10^{-6}$, the Gauss-Bonnet model with monomial coupling can successfully describe both the preheating process and gravitational wave production.

\end{document}